\PassOptionsToPackage{table}{xcolor}
\documentclass[sigconf]{acmart}

\AtBeginDocument{
  }

\setcopyright{acmlicensed}
\copyrightyear{2018}
\acmYear{2018}
\acmDOI{XXXXXXX.XXXXXXX}

\acmConference[33rd ACM International Conference on Multimedia]{Make sure to enter the correct
  conference title from your rights confirmation email}{October 27--31,
  2025}{Dublin, Ireland}

\acmISBN{978-1-4503-XXXX-X/2018/06}
\acmSubmissionID{3502}
\usepackage{caption}  
\usepackage{multirow}
\usepackage{makecell}
\usepackage{etoolbox}

\usepackage{float}
\usepackage{booktabs}
\usepackage{float}
\definecolor{lightgray}{gray}{0.92}
\definecolor{midgray}{gray}{0.85}
\definecolor{darkgray}{gray}{0.78}
\author{Jiaxun Zhang}
\affiliation{
  \institution{State Key Lab of IoT for Smart City, Dept. of Civil and Environmental Engineering, University of Macau}
  \city{Macau SAR}
  \country{China}
}

\author{Haicheng Liao}
\affiliation{
  \institution{State Key Lab of IoT for Smart City, Dept. of Computer and Information Science, University of Macau}
  \city{Macau SAR}
  \country{China}
}

\author{Yumu Xie}
\affiliation{
  \institution{State Key Lab of IoT for Smart City, Dept. of Computer and Information Science, University of Macau}
  \city{Macau SAR}
  \country{China}
}

\author{Chengyue Wang}
\affiliation{
  \institution{State Key Lab of IoT for Smart City, Dept. of Civil and Environmental Engineering, University of Macau}
  \city{Macau SAR}
  \country{China}
}

\author{Yanchen Guan}
\affiliation{
  \institution{State Key Lab of IoT for Smart City, Dept. of Civil and Environmental Engineering, University of Macau}
  \city{Macau SAR}
  \country{China}
}

\author{Bin Rao}
\affiliation{
  \institution{State Key Lab of IoT for Smart City, Dept. of Civil and Environmental Engineering, University of Macau}
  \city{Macau SAR}
  \country{China}
}

\author{Zhenning Li}
\orcid{0000-0002-0877-6829}
\authornote{Corresponding author.} 
\affiliation{
  \institution{State Key Lab of IoT for Smart City, Depts. of Civil and Environmental Engineering and Computer and Information Science, University of Macau}
  \city{Macau SAR}
  \country{China}
}
\email{zhenningli@um.edu.mo}       

\begin{document}

\let\WriteBookmarks\relax
\def\floatpagepagefraction{1}
\def\textpagefraction{.001}

\title[CAMERA: Context-Aware Multi-modal Enhanced Risk Anticipation]{Eyes on the Road, Mind Beyond Vision: Context-Aware Multi-modal Enhanced Risk Anticipation}

\renewcommand{\shortauthors}{Zhang et al.}

\begin{abstract}
Accurate accident anticipation remains challenging when driver cognition and dynamic road conditions are underrepresented in predictive models. In this paper, we propose \textbf{CAMERA} (\emph{Context-Aware Multi-modal Enhanced Risk Anticipation}), a multi-modal framework integrating dashcam video, textual annotations, and driver attention maps for robust accident anticipation. Unlike existing methods that rely on static or environment-centric thresholds, CAMERA employs an adaptive mechanism guided by scene complexity and gaze entropy, reducing false alarms while maintaining high recall in dynamic, multi-agent traffic scenarios. A hierarchical fusion pipeline with Bi-GRU (Bidirectional GRU) captures spatio-temporal dependencies, while a Geo-Context Vision-Language module translates 3D spatial relationships into interpretable, human-centric alerts. Evaluations on the DADA-2000 and benchmarks show that CAMERA achieves state-of-the-art performance, improving accuracy and lead time. These results demonstrate the effectiveness of modeling driver attention, contextual description, and adaptive risk thresholds to enable more reliable accident anticipation. 
\end{abstract}



\renewcommand\footnotetextcopyrightpermission[1]{}
\settopmatter{printacmref=false}
\begin{teaserfigure}
  \includegraphics[width=\textwidth]{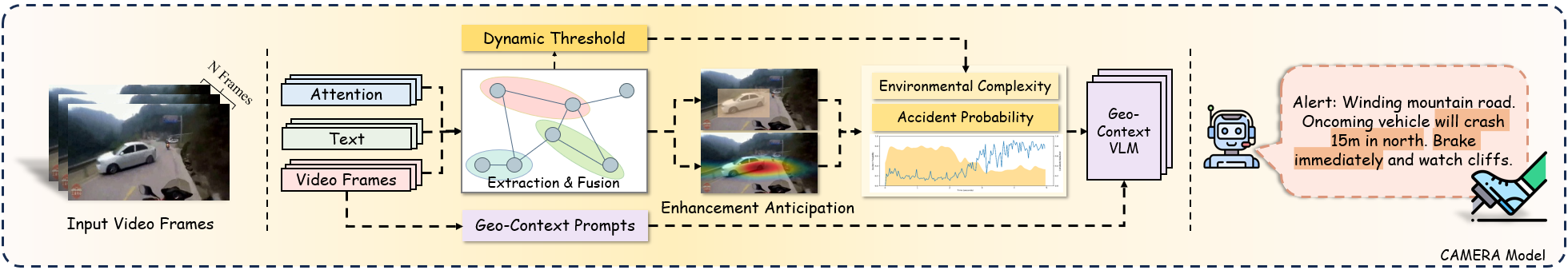}
    \caption{\textbf{Overview of CAMERA's anticipation process.} CAMERA fuses video, text, and attention maps to predict accident probability and a dynamic risk threshold, while a Geo-Context VLM generates alerts to inform passengers of potential dangers.}
  \label{intro}
\end{teaserfigure}
\maketitle

\section{Introduction}

Traffic accidents persist globally, resulting in over 1.35 million annual fatalities and significant economic impacts \citep{world2023global,NHTSA2023}. Enabled by growing sensor data and dashcam repositories, studies employ deep learning for real-time high-risk scenario prediction, aiming to reduce accidents through proactive driver assistance \citep{beck2023automated, lin2024near, cai2025v}. However, accurate accident identification and anticipation across diverse road conditions, complex interactions, and human behavioral variations continue to pose critical research challenges.

Current accident-anticipation research prioritizes integrating multimodal data effectively. Early approaches focused on vision-centric cues—dashcam footage, vehicle/pedestrian tracking, and trajectory-based forecasting \citep{verma2024vision, papadopoulos2024lightweight}. Modern methods incorporate driver-centric factors like gaze patterns and cognitive states, recognizing human attention lapses as key accident contributors \citep{zeng2017agent, lachance2025gaze}. Contextual sources like textual road event annotations and weather data further enhance risk assessment \citep{xie2025driving, zhang2025latte}. Yet unifying visual, textual, and driver-related streams remains challenging, as simplistic fusion strategies often neglect critical inter-modal interactions.

Another longstanding limitation in accident anticipation arises from \emph{static thresholding}, where a fixed confidence level triggers warnings across all scenarios. Such an approach risks frequent false alarms in benign environments (e.g., highways) or failing to detect threats in denser traffic (e.g., urban intersections) \citep{chowdhury2023flamingo,liu2025vlm,liao2024gpt}. Meanwhile, \emph{Vision-Language Models (VLMs)} have emerged as powerful tools for explaining visual phenomena in natural language, enabling interpretable system outputs \citep{radford2021learning,grigsbyvlm}. Yet, many VLM-based accident anticipation systems still face difficulties integrating \emph{3D spatial reasoning}, leading to potential misjudgments of distance, occlusion, and object overlap under driving complexities \citep{wandelt2024large,mahmud2025integrating}.

To address these challenges, we propose \textbf{CAMERA}—\emph{Context-Aware Multi-modal Enhanced Risk Anticipation}—a \emph{cognitively informed} unified framework that harmonizes three core perspectives to predict imminent accidents more reliably:
\begin{enumerate}
    \item \textbf{Dashcam Vision and Textual Context:} CAMERA encodes video frames alongside textual annotations describing road events, weather conditions, or scene context, capturing both spatial and semantic dimensions of risk.
    \item \textbf{Driver Attention States:} By integrating gaze and attention heatmaps, the system learns how driver focus—or lack thereof—affects accident likelihood, addressing a vital human factor often underrepresented in traditional models.
    \item \textbf{Adaptive Thresholding:} A novel \emph{scene complexity metric} dynamically modulates the warning threshold, informed by both environmental entropy (e.g., traffic density) and driver attention uncertainty. This approach avoids the pitfalls of fixed confidence levels, providing earlier and more accurate predictions under challenging conditions.
\end{enumerate}

Moreover, CAMERA employs a \emph{Geo-Context VLM} component that grounds risks in 3D spatial references (e.g., “Pedestrian entering left blind spot”), offering interpretable alerts aligned with driver perception. By synchronously modeling visual dynamics, contextual signals, and driver cognition, CAMERA represents a next-generation, human-aligned solution for proactive traffic safety.

We systematically evaluate CAMERA on
\emph{DADA-2000} and other three \emph{benchmarks}—demonstrating its ability to outperform conventional baselines in both prediction accuracy and early-warning capacity. In particular, we find that CAMERA yields a markedly lower false-positive rate while simultaneously identifying accidents at longer lead times compared to static-threshold systems. In summary, our contributions include the following:  
\begin{itemize}
    \item We introduce a hierarchical architecture unifying video frames, driver attention maps, and textual annotations via a synchronized feature fusion mechanism, capturing robust spatio-temporal cues for accident anticipation.
    \item By incorporating a data-driven scene complexity metric, CAMERA adapts risk sensitivity based on attention entropy and textual embedding complexity, offering context-aware warnings that significantly reduce false positives.
    \item Geo-Context VLM translates numerical spatial relationships into interpretable, actionable alerts, bridging technical hazard prediction with intuitive driver communication.
    \item Evaluations on benchmark datasets show consistent gains in accuracy and anticipation time over fixed-threshold baselines, highlighting CAMERA’s practicality for real-world deployment in Advanced Driver-Assistance Systems (ADAS).
\end{itemize}

\section{Related Work}
Preventing accidents proactively is an enduring goal in intelligent transportation, as early warning can help prevent serious consequences. Initially, researchers focused on dashcam-based spatio-temporal cue modeling using trajectory analysis and object tracking \citep{liao2024crash,choudhury2025intelligent}. Graph Convolutional Networks (GCNs) enhanced scenario understanding and reduced false positives by representing vehicles and road agents as graph nodes and modeling their interactions over time in controlled settings \citep{gan2025graph,chai2025gacnet,liao2025minds}. While effective at identifying vehicle-centric risks, these methods often overlook driver-specific factors, such as distraction or intent, which contribute to various kinds of serious traffic incidents \citep{Karim2021stdan, liao2024cognitive, banisalman2025vrdeepsafety,huo20253d}. 

Recognizing the pivotal role of human behaviors, recent work has incorporated physiological signals, saliency maps, or gaze tracking to capture driver inattention \citep{monjurul2021towards,hussain2025empirical,weibull2025driver}. Studies involving hierarchical fusion of spatio-temporal features demonstrated improved performance on benchmark dashcam datasets, emphasizing that understanding \emph{how and where} drivers focus could enhance anticipation of near-accident moments. However, many of these approaches analyze driver states in isolation rather than synthesizing them with dynamic road contexts and scene-level data. Consequently, they lack adaptive mechanisms to adjust risk thresholds when traffic density spikes or unexpected driving patterns emerge.

In parallel, multi-modal accident-anticipation frameworks have gained traction. By integrating RGB video with textual context (e.g., annotated descriptions of weather or road events) or GPS signals, researchers achieved more robust predictions and more interpretable outputs. VLMs further advanced interpretability by automatically generating textual risk assessments from visual inputs \citep{chowdhury2023flamingo,shi2025scvlm,teng5112947improving}. Despite these promising developments, standard VLM pipelines often struggle with precise 3D spatial inference, leading to errors in distance estimation or object-scale perception when traffic scenarios evolve rapidly \citep{zhang2025latte, jain2024semantic,liao2025cot}. Additionally, relying on expensive sensors like LiDAR raises deployment concerns under adverse weather or cost-sensitive settings. Moreover, current VLM-based solutions seldom incorporate driver-centric attention signals or adapt their risk-detection thresholds in real time, potentially undermining usability in complex, real-world conditions.

Collectively, the literature illustrates that accurate and timely accident anticipation hinges on two major requirements: (1) unifying driver attention with environmental context in a single predictive pipeline, and (2) dynamically tuning risk alerts to reflect shifting scene complexity and cognitive load. To meet these challenges, our approach integrates multi-modal data—RGB dashcam frames, textual descriptions, and driver attention patterns—within a framework that dynamically adapts its thresholds based on a learned complexity metric. In addition, we adopt a 3D-aware vision-language module to provide interpretable, spatially grounded alerts, enabling accident anticipation that is not only more accurate but also more aligned with driver perceptions and practical real-time assistance.

\begin{figure*}[htbp]
    \centering
    \includegraphics[width=0.95\textwidth]{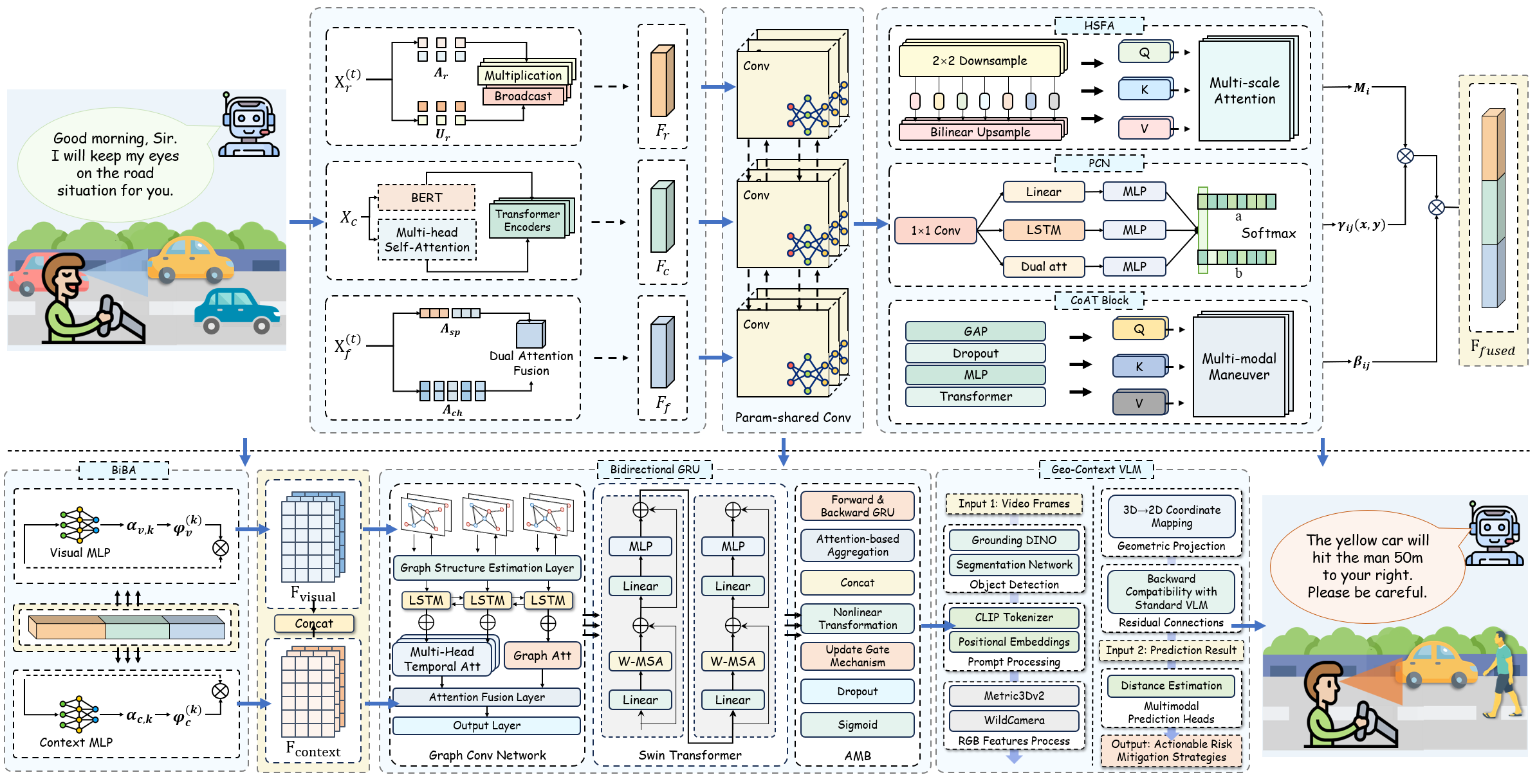}
        \vspace{-8pt}
\caption{Overall architecture of our proposed CAMERA model, illustrating the pipeline from multi-modal inputs (RGB frames, scene context, driver attention) to risk anticipation and interpretable geo-context alerts. Modules include HSFA (Hierarchical Scale Fusion with Attention), PCN (Position-wise Correspondence Network), CoAT (Cross-Modal Co-Activation Transformer), BiBA (Bifurcated Basis Aggregation), and AMB (Adaptive Memory Blending).}
    \label{fig:enter-label1}
\end{figure*}

\section{Methodology}
\subsection{Problem Formulation}

The goal of the CAMERA framework is to predict traffic accidents and generate interpretable risk explanations using multi-modal data. Given an input dashcam video sequence \( \mathcal{V} = \{ v_t \}_{t=1}^T \) consisting of \( T \) frames, our model integrates three complementary modalities: (1) RGB video frames \( \mathbf{X}_r^{(t)} \), encoding visual scene information; (2) textual descriptions \( \mathbf{X}_c \), providing contextual semantics (e.g., weather, road layout); and (3) driver attention maps \( \mathbf{X}_f^{(t)} \), indicating real-time gaze and focus patterns. The primary predictive tasks are formulated as: (i) estimating the probability of an accident occurring at each frame, \( p_t \in [0,1] \); (ii) generating spatially precise risk maps \( \mathbf{M}_s^{(t)} \); and (iii) triggering proactive alerts when the predicted risk surpasses an adaptive threshold, i.e., \( p_t > \tau_t \), with \( \tau_t \in [0.3, 0.7] \).

Unlike traditional approaches employing fixed risk thresholds \citep{zhang2025traffic,liao2024real}, CAMERA introduces a novel dynamic thresholding mechanism that adjusts sensitivity based on real-time scene complexity derived from contextual semantics \( \mathbf{X}_c \). Specifically, this adaptive threshold is computed using a combination of driver cognitive load indicators and scene entropy measures, thus effectively reducing false alarms and maintaining high accuracy in challenging scenarios such as low-visibility conditions or congested intersections. To ensure accurate predictions, CAMERA jointly optimizes a hybrid loss function, employing focal loss to handle classification imbalance and KL divergence to align predicted risk maps with driver attention distributions. A Bi-GRU-based temporal model explicitly captures accident precursor signals within a critical anticipation window of 3–5 seconds, matching typical human reaction capabilities \citep{kumamoto2025aat,kim2025understanding,guan2025experimental}. When risk exceeds the adaptive threshold (\( p_t > \tau_t \)), CAMERA utilizes its Geo-Context VLM module to produce human-centric alerts detailing specific risk types (e.g., pedestrian crossing) and spatial references (e.g., "northeast"), enhancing real-world applicability in various dynamic traffic scenarios \citep{abdulrashid2024explainable,munir2025pedestrian,guo2024vlm}.

\subsection{Framework Overview}

CAMERA offers an end-to-end solution for accident anticipation and interpretation through a structured four-stage pipeline, illustrated in Fig.~\ref{fig:enter-label1}. Initially, \textbf{Multi-Modal Feature Extraction} processes RGB dashcam sequences through channel-wise calibration, refines driver attention maps using multi-scale spatial attention, and extracts semantic embeddings from textual descriptions via a transformer-based module. These modality-specific representations are then passed to the \textbf{Adaptive Hierarchical Fusion} stage, which employs hierarchical spatial transformers and task-aware decomposition strategies to harmonize multi-modal information.

Subsequently, compressed and fused features feed into a \textbf{Bidirectional GRU Network} to capture complex temporal dependencies and anticipate critical precursor events leading to accidents. Temporal modeling is enhanced by an adaptive memory mechanism, allowing the network to effectively track evolving risk indicators. The final \textbf{CAMERA Prediction} stage synthesizes spatiotemporal cues, dynamically adjusts prediction thresholds based on scene complexity metrics and driver attention entropy, and outputs both frame-level accident probabilities and spatially localized risk maps.

Additionally, the integrated \textbf{Geo-Context VLM} converts model predictions into human-centric verbal alerts by mapping risk estimates to explicit 3D spatial coordinates and generating natural language descriptions, enhancing interpretability and usability. Trained via multi-loss optimization balancing temporal coherence, modality alignment, and interpretability, CAMERA framework delivers robust, context-aware, and interpretable accident anticipation suitable for real-world traffic safety scenarios.

\subsection{Multi-Modal Feature Extraction}

The multi-modal feature extraction module processes RGB video sequences, textual descriptions, and driver attention maps through specialized pathways, resulting in modality-specific embeddings aligned for downstream fusion. For RGB frames \(\mathbf{X}_r^{(t)} \in \mathbb{R}^{H \times W \times 3}\), we extract convolutional feature maps \(\mathbf{U}_r = f_{\text{conv}}(\mathbf{X}_r^{(t)})\) using a standard backbone network. These features are designed to preserve essential visual semantics (e.g., vehicles, pedestrians) and maintain robustness against challenging traffic conditions. The resulting feature representations \(\mathbf{F}_r \in \mathbb{R}^{h \times w \times c}\) are then forwarded to the fusion module together with features from other modalities.

In parallel, contextual textual inputs \( \mathbf{X}_c \) (e.g., traffic descriptions such as “vehicle running red light”) are processed by a transformer-based encoder. Using a pre-trained BERT-base model, textual embeddings \(\mathbf{F}_c \in \mathbb{R}^{h \times w \times c}\) are derived from the  BERT's tokens after processing through a 12-layer transformer encoder with multi-head self-attention. These embeddings succinctly encode semantic context, enabling subsequent cross-modal integration.

To enhance critical information in driver attention maps \(\mathbf{X}_f^{(t)} \in \mathbb{R}^{H \times W}\), we adopt a dual-stage attention mechanism comprising channel-wise and spatial refinement. First, channel-wise attention is derived from global average and max pooling descriptors, which are passed through fully connected layers to generate attention weights that reflect each channel's relevance to driving semantics. The channel-refined features are then downsampled via convolution to a fixed spatial resolution. Next, spatial attention is computed by aggregating channel-wise average and max descriptors, followed by a convolution to emphasize gaze-focused regions. The final attentional feature representation is obtained as follows:
\begin{equation}
\mathbf{F}_f = \mathbf{A}_{sp} \odot \left(\mathbf{A}_{ch} \circledast \mathbf{X}_f^{(t)}\right)
\end{equation}
where \(\circledast\) denotes convolution and \(\odot\) represents element-wise multiplication. To preserve both fine-grained and coarse-level gaze patterns, a 4-level spatial pyramid is constructed using strided convolutions. Each level undergoes independent attention refinement, and scale-specific weights are fused via softmax-weighted upsampling. This structure suppresses irrelevant background while amplifying risk-aware regions. The resulting features \(\mathbf{F}_f \in \mathbb{R}^{h \times w \times c}\) are dimensionally aligned with other features, facilitating effective cross-modal fusion enriched with gaze-informed semantic priors.

\subsection{Adaptive Hierarchical Fusion}

The extracted multi-modal features—visual (\(\mathbf{F}_r\)), textual (\(\mathbf{F}_c\)), and driver attention (\(\mathbf{F}_f\))—are integrated into a unified representation via an adaptive hierarchical fusion mechanism. Each modality is first projected into a common feature space using dynamic convolutional layers with shared-filter kernels, producing aligned feature representations. Hierarchical multi-resolution structures are then constructed through designed downsampling and upsampling operations, facilitating rich cross-scale interactions across modalities.

To capture complementary information across multiple scales, each modality forms a spatial pyramid representation with multi-resolution features. HSFA module dynamically weighs these scales via learnable attention, recalibrating each modality’s representation into a scale-aware feature \(\mathbf{M}_i\). Cross-modal interaction is then performed using a dual attention mechanism that models inter-modal dependencies both at the channel and spatial levels. Channel-wise co-activation weights and spatial correspondence maps guide feature alignment and integration. The resulting fused representation is formulated as follows:
\begin{equation}
\mathbf{F}_{\text{fused}} = \sum_{i,j} \beta_{ij} \odot \left(\sum_{x,y} \gamma_{ij}(x,y) \cdot \mathbf{H}_i(x,y)\right),
\end{equation}
where \(\beta_{ij}\) dynamically adjusts feature emphasis between modality pairs \((i,j)\), and \(\gamma_{ij}(x,y)\) captures spatial dependencies via cosine similarity. This calibration ensures adaptive alignment of complementary information from RGB, textual, and attention streams.

Subsequently, we perform task-specific decomposition of \(\mathbf{F}_{\text{fused}}\) into visual-centric (\(\mathbf{F}_{\text{visual}}\)) and context-centric (\(\mathbf{F}_{\text{context}}\)) representations. Employing modality-specific basis projections \(\mathbf{\Phi}_v^{(k)}, \mathbf{\Phi}_c^{(k)}\), we achieve decomposition as follows:
\begin{equation}
\mathbf{F}_{\text{visual}} = \sum_{k=1}^{K} \alpha_{v,k} \cdot \mathbf{\Phi}_v^{(k)} \mathbf{F}_{\text{fused}}, \quad
\mathbf{F}_{\text{context}} = \sum_{k=1}^{K} \alpha_{c,k} \cdot \mathbf{\Phi}_c^{(k)} \mathbf{F}_{\text{fused}},
\end{equation}
where \(\alpha_{v,k}\) and \(\alpha_{c,k}\) are adaptive coefficients computed through global pooling and softmax-normalized perceptrons. This bifurcation enables specialized processing: visual-centric features emphasize spatial details vital for accident localization, while context-centric features focus on semantic interpretations required for accurate accident classification and explanation. Thus, our framework ensures coherent, interpretable, and robust representations through hierarchical fusion and adaptive calibration, critical for accurate, timely, and actionable accident anticipation.

\subsection{Bidirectional GRU Network}

To effectively capture temporal dependencies critical for accident anticipation, the proposed framework employs a Bi-GRU network. The Bi-GRU processes fused multi-modal features, which include visual representations (\(\mathbf{F}_{\text{visual}}^{(t)}\)) and contextual semantics (\(\mathbf{F}_{\text{context}}^{(t)}\)). Concatenated into \(\mathbf{F}_{\text{in}}^{(t)} = [\mathbf{F}_{\text{visual}}^{(t)}; \mathbf{F}_{\text{context}}^{(t)}] \in \mathbb{R}^{h \times w \times 2c}\), these inputs are processed. To reduce computational complexity while retaining essential spatial and semantic information, we apply spatial average pooling, yielding a compact feature vector \(\mathbf{\overline{F}}_{\text{in}}^{(t)} \in \mathbb{R}^{2c}\).

The Bi-GRU captures sequential context by computing forward (\(\overrightarrow{\mathbf{h}}_t\)) and backward (\(\overleftarrow{\mathbf{h}}_t\)) hidden states. Each direction employs gating mechanisms—update gate (\(\mathbf{z}_t\)) and reset gate (\(\mathbf{r}_t\))—to dynamically balance historical and current observations:
\begin{equation}
\mathbf{z}_t = \mu\left(\mathbf{W}_z[\mathbf{h}_{t-1}; \mathbf{\overline{F}}_{\text{in}}^{(t)}] + \mathbf{b}_z\right), \quad
\mathbf{r}_t = \mu\left(\mathbf{W}_r[\mathbf{h}_{t-1}; \mathbf{\overline{F}}_{\text{in}}^{(t)}] + \mathbf{b}_r\right),
\end{equation}
where \(\mu\) denotes the Swin Transformer, \( [\cdot;\cdot] \) represents graph convolution network., and \(\mathbf{W}_z, \mathbf{W}_r, \mathbf{b}_z, \mathbf{b}_r\) are learnable parameters. Subsequently, a candidate hidden state \(\mathbf{\tilde{h}}_t\) is computed using an adaptive memory blending (AMB) mechanism, which integrates historical context modulated by \(\mathbf{r}_t\). Formally,
\begin{equation}
\mathbf{\tilde{h}}_t = \text{tanh}\left(\mathbf{W}_h[\mathbf{r}_t \odot \mathbf{h}_{t-1}; \mathbf{\overline{F}}_{\text{in}}^{(t)}] + \mathbf{b}_h\right).
\end{equation}
where \(\odot\) denotes element-wise multiplication. The final hidden state is updated as a weighted combination of past and present:
\begin{equation}
\mathbf{h}_t = (1 - \mathbf{z}_t) \odot \mathbf{h}_{t-1} + \mathbf{z}_t \odot \mathbf{\tilde{h}}_t.
\end{equation}
Forward and backward states are concatenated to form comprehensive temporal representations: \(\mathbf{H}_t^{\text{gru}} = [\overrightarrow{\mathbf{h}}_t; \overleftarrow{\mathbf{h}}_t]\). By modeling both past and future context within a 3–5 second window—aligned with typical human reaction times \citep{anjum2023spatio}—the Bi-GRU network robustly encodes critical accident precursor signals.

\begin{figure*}
    \centering
    \vspace{-5pt}
    \includegraphics[width=1\textwidth]{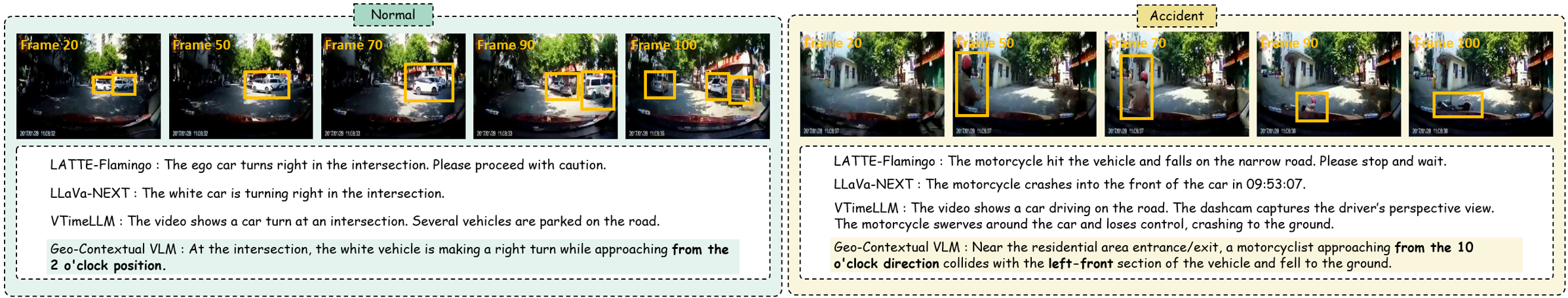}
    \vspace{-15pt}
    \caption{Comparison of Geo-Context VLM with LATTE-Flamingo\citep{zhang2025latte}, LLaVa-NEXT\citep{liao2024and} and VTimeLLM\citep{huang2024vtimellm} in different situation. }
    \label{fig:enter-label6}
\end{figure*}
\subsection{Geo-Context Vision-Language Model}

While conventional VLMs primarily facilitate general image-text matching or simple descriptive tasks, our Geo-Context VLM introduces robust spatial awareness and proactive accident anticipation, significantly extending typical VLM capabilities (Fig.~\ref{fig:enter-label6}). The Geo-Context VLM consists of three integrated components: object detection and segmentation, spatially aligned textual prompting, and metric-aware visual-language fusion.

Object detection adopts a dual-stage process: localizing objects with Grounding Dino \citep{liu2024grounding,devanna2025boosting} and refining detection masks with segmentation networks \citep{ke2023segment,li2025visual} for pixel-accurate localization. Textual prompts combining qualitative ("left") and quantitative ("3m") spatial cues undergo CLIP-based tokenization with positional encodings. Special tokens dynamically link textual and visual features.

For robust 3D spatial inference, Geo-Context VLM integrates Metric3Dv2 \citep{mahmoudieh2022zero,hayes2025revisiting} for accurate depth estimation and WildCamera \citep{marino2019ok,zhu2023tame} for camera calibration, enabling precise metric predictions. By leveraging residual connections, these metric-based embeddings seamlessly integrate into traditional VLM pipelines. Consequently, Consequently, Geo-Context VLM synthesizes visual frames, real-time predictive scores (e.g., accident probability, mTTA) from the CAMERA training process module to generate interpretable and spatially precise accident alerts ("Pedestrian 2.1m ahead in left blind spot"), which can enhance safety in dynamic driving scenes.  

\subsection{Training}
This module estimates accident probability (\(p_t \in [0,1]\)) by synthesizing temporal and contextual cues. Specifically, Bi-GRU outputs (\(\mathbf{H}_t^{\text{gru}} \in \mathbb{R}^{2d}\)) are integrated with context-based complexity features extracted from textual descriptors (\(\mathbf{F}_{\text{context}}^{(t)}\)). Context features undergo nonlinear transformation via a two-layer MLP (\(\psi(\cdot)\)) equipped with ReLU activation, yielding complexity-aware representations. The final prediction is computed as:
\begin{equation}
p_t = \sigma\left(\mathbf{w}_p^\top [\mathbf{H}_t^{\text{gru}} \oplus \psi(\mathbf{F}_{\text{context}}^{(t)})] + b_p\right),
\end{equation}
where \(\oplus\) denotes vector concatenation, and parameters \(\mathbf{w}_p, b_p\) are learned end-to-end. For spatial risk localization, upsampled GRU states (\(\mathbf{\tilde{H}}_t^{\text{gru}} \in \mathbb{R}^{h \times w \times d}\)) are combined with visual features (\(\mathbf{F}_{\text{visual}}^{(t)}\)) through channel-wise cross-correlation. The resulting spatial attention map (\(\mathbf{M}_s^{(t)}\)) emphasizes areas with elevated accident risks:
\begin{equation}
\mathbf{M}_s^{(t)} = \mathcal{S}\left(\sum_{k=1}^d \mathbf{\tilde{H}}_t^{\text{gru}}[:,:,k] \circledast \mathbf{F}_{\text{visual}}^{(t)}[:,:,k]\right),
\end{equation}
with \(\mathcal{S}\) denoting the softmax normalization. To reduce false alarms and increase robustness, the threshold \(\tau_t\) dynamically modulates the prediction sensitivity while \( \lambda_1, \lambda_2 \in [0,0.2] \) are learnable coefficients ensuring \(\tau_t \in [0.3, 0.7]\), based on driver attention uncertainty (\(\mathcal{E}(\mathbf{X}_f^{(t)})\)) and scene complexity (\(\|\mathbf{F}_{\text{context}}^{(t)}\|_2\)):  
\begin{equation}
\tau_t = 0.5 + \lambda_1 \mathcal{E}(\mathbf{X}_f^{(t)}) - \lambda_2 \|\mathbf{F}_{\text{context}}^{(t)}\|_2.
\end{equation}
The training objective optimizes a composite loss \(\mathcal{L}\), integrating focal loss to handle prediction imbalance, KL divergence for aligning spatial risk maps with driver attention, and temporal smoothness loss to penalize abrupt fluctuations:
\begin{equation}
\mathcal{L} = \mathcal{L}_{\text{focal}} + 0.5 \mathcal{L}_{\text{KL}} + 0.1 \mathcal{L}_{\text{smooth}}.
\end{equation}
\section{Experiment}
\subsection{ Datasets }
We evaluate CAMERA using four publicly available datasets:
\begin{itemize}
    \item \textbf{DADA-2000:} The Driver Attention in Driving Accident Scenarios-2000 (DADA-2000) \citep{fang2021dada} provides driver attention analysis via 2,000 videos (658k frames),  integrating crowd/lab tracking with 54 crash types. It aligns crash annotations (1584×660) with gaze metrics while preserving pre-crash sequences for attention mechanism analysis.
    \item \textbf{CCD:} The Car Crash Dataset (CCD) \citep{bao2020uncertainty} documents environmental factors, ego-vehicle dynamics, involved parties, and accident causality. Comprising 1,500 positive and 3,000 negative clips (50-frame/5s each), the dataset splits into 3,600 training and 900 testing samples with 4:1 ratio.
    \item \textbf{DAD:}The Dashcam Accident Dataset (DAD) \citep{chan2017anticipating} contains 720p dashcam footage from six Taiwanese cities, with 620 positive and 1,130 negative clips divided into 1,284 training and 466 testing clips (100 frames/5 seconds each). 
    \item \textbf{A3D:} The AnAn Accident Detection Dataset (A3D) \citep{yao2019unsupervised} captures abnormal road events in East Asian urban environments, comprising 1,087 positive and 114 negative clips split into 961 training and 240 testing samples. 
\end{itemize}

\subsection{Evaluation Metrics}
To evaluate CAMERA’s accident anticipation performance, we adopt four widely used metrics:

\begin{itemize}
    \item \textbf{Average Precision:} AP quantifies the precision-recall curve by integrating precision values across all recall levels, with higher values indicating more reliable performance.

    \item \textbf{Area Under Curve:} AUC measures the area under the ROC curve, representing the probability that a positive sample is ranked higher than a negative one.

    \item \textbf{Time-To-Accident at Recall 50\%:} TTA@R50 captures the average time difference between predicted and actual accident points when the model achieves 50\% recall.

    \item \textbf{Mean Time-To-Accident:} mTTA measures the average anticipatory lead time in correctly anticipated accidents.
\end{itemize}

\subsection{Implementation Details}

Experiments used two NVIDIA A6000 GPUs (48GB VRAM). The PyTorch 1.13/Python 3.9-implemented CAMERA framework with AdamW optimizer (weight decay 0.01) and cosine learning scheduling. Learning rates underwent linear warm-up to \(1\times10^{-4}\) in initial 10 epochs, decaying to \(1\times10^{-5}\). Peak rates were \(1\times10^{-3}\) for fusion/GRU modules and \(1\times10^{-4}\) for Geo-Context VLM. Batch size 2 accommodated memory-intensive multi-modal fusion and spatio-temporal modeling. Temporal sampling employed 5-second sliding windows. RGB frames resized to \(224\times224\) with attention maps downsampled to visual feature resolution. Text inputs tokenized via BERT-base, truncated/padded to 32 tokens. Training augmentations included RGB brightness/contrast shifts, attention map Gaussian blur, and text dropout masking. The loss function incorporated focal loss, KL-divergence alignment, and temporal smoothness regularization. Gradient clipping (1.0) and validation AP-based early stopping ensured stability. Streamlined inference ran at 5 FPS with adaptive thresholds based on scene complexity/attention Sentropy. After backbone freezing, Geo-Context VLM was fine-tuned for 30 epochs (\(5\times10^{-5}\) lr). 

\subsection{Comparison to State-of-the-art Baselines}

We compare CAMERA with DRIVE \citep{bao2021drive} and LOTVS-CAP \citep{li2024cognitive} on DADA-2000 accident anticipation under identical protocols. As Table~\ref{tab:my_label1} shows, CAMERA sets new DADA-2000 benchmarks with 80.52\% AP and 89.64\% AUC, significantly outperforming prior methods. Our framework achieves \textbf{+5.5\% AP} and \textbf{+7.6\% AUC} gains over LOTVS-CAP \citep{li2024cognitive}, while maintaining superior temporal anticipation (4.0583s TTA@R50, 4.4837s mTTA). CAMERA shows \textbf{16.4\% higher AP} than DRIVE \citep{bao2021drive}, with \textbf{29.9\% greater AUC}, demonstrating enhanced discriminative power for safety-critical anticipation. While LOTVS-CAP shows competent temporal awareness (3.95s TTA@R50), CAMERA extends leads via dynamic complexity reduction, achieving \textbf{2.7\% longer anticipation windows} under stricter recall constraints. These advances confirm that integrating driver attention with multi-modal spatiotemporal reasoning improves anticipation accuracy and early warnings in complex traffic.
\begin{table}[H]
    \centering
    \vspace{-8pt} 
    \scalebox{0.7}{
    \resizebox{0.6\textwidth}{!}{
    \begin{tabular}{c|c|c|c|c}
        \hline
         & AP & AUC & TTA@50 & mTTA \\
        \hline
        DRIVE \citep{bao2021drive} & 0.72 & 0.69 & 3.657 & 4.295 \\
        LOTVS-CAP \citep{li2024cognitive} & \underline{0.75} & \underline{0.82} & \underline{3.951} & \underline{4.464} \\
        \textbf{CAMERA} & \textbf{0.8052} & \textbf{0.8964} & \textbf{4.0583} & \textbf{4.4837} \\
        \hline
    \end{tabular}
    }}
    \caption{Comparison of models for the highest metrics on DADA-2000. \textbf{Bold} and \underline{underlined} denote top and second-best results.}
    \label{tab:my_label1}
    \vspace{-20pt}
\end{table}

Cross-dataset validation demonstrates CAMERA’s robust accident anticipation across diverse scenarios. On CCD’s accident-focused benchmark, CAMERA achieves state-leading 99.7\% AP (matching W3AL \citep{liao2024and}) with superior 4.76s mTTA, outperforming trajectory-based GSC \citep{wang2023gsc} by +1.18s in temporal anticipation. Crucially, CAMERA maintains high reliability in complex real-world scenes: 77.8\% AP on DAD’s mixed urban scenarios, surpassing W3AL by +8.6\% AP while delivering actionable 4.18s mTTA warnings. For A3D’s culturally distinct East Asian events, CAMERA attains 97.1\% AP – +1.1\% over CRASH \citep{liao2024crash} – with near-optimal 4.74s mTTA, significantly outperforming GSC’s 2.12s latency. This tri-dataset consistency reveals CAMERA’s unique balance of precision and timeliness: >99.7\% AP in controlled extremes (CCD) versus 77.8-97.1\% AP in naturalistic settings (DAD/A3D), coupled with consistently safe mTTA (>4.18s across all benchmarks). The model’s cross-scenario adaptability – handling accident dynamics, dense urban interactions, and regional anomalies – confirms its generalization beyond single-dataset optimization. Such multi-environment reliability, evidenced by +8.6\% AP gains on DAD’s mixed traffic and +2.12s mTTA advantage over GSC in early anomaly prediction, positions CAMERA as a robust solution for ADAS deployment.

\begin{table}[h]
\vspace{-5pt}
    \centering
    \label{tab:my_label2}
    \scalebox{0.75}{ 
    \begin{tabular}{c|c|c|c|c|c|c}
        \hline
        & \multicolumn{2}{c|}{CCD} & \multicolumn{2}{c|}{DAD} & \multicolumn{2}{c}{A3D} \\
        \cline{2-7}
        & AP (\%) & mTTA & AP (\%) & mTTA & AP (\%) & mTTA \\
        \hline
        Ustring \citep{bao2020uncertainty} & 99.5 & 4.73 & 53.7 & 3.53 & 94.4 & \textbf{4.92} \\
        GSC \citep{wang2023gsc} & 99.3 & 3.58 & 60.4 & 2.55 & 94.9 & 2.62 \\
        CRASH \citep{liao2024crash} & \underline{99.6} & \textbf{4.91} & 65.3 & 3.05 & \underline{96.0} & \textbf{4.92} \\
        W3AL \citep{liao2024and} & \textbf{99.7} & 3.93 & \underline{69.2} & \textbf{4.26} & 92.41 & 4.52 \\
        CAMERA & \textbf{99.7} & \underline{4.76} & \textbf{77.8} & \underline{4.18} & \textbf{97.1} & \underline{4.74} \\
        \hline
    \end{tabular}
    }
    \caption{Comparisons for the highest metrics on three benchmarks. \textbf{Bold} and \underline{underlined} denote top and second-best results.}
    \vspace{-18pt}
\end{table}

\subsection{Ablation Studies}
Ablation studies are crucial in accident anticipation models, as they test whether the model can distinguish true causal precursors from spurious correlations in the data. Through systematic training degradation, Table~\ref{tab:my_label3} demonstrates learning prioritization: maintained accuracy with missing data suggests physics-grounded reasoning, while performance collapse indicates overfitting to spurious patterns absent in information-limited environments.

Table~\ref{tab:my_label3} showcases CAMERA's data-efficient accident anticipation across completeness levels, achieving AP gains of +0.07\% (50\%), +0.6\% (75\%), and +5.52\% (100\%) over second-place LOTVS-CAP. Its event-context disentanglement enables superior label efficiency via linear scaling: retaining 92.1\% full-data AP (0.7413 vs. 0.8052) at 50\% data versus DRIVE's 91.2\% (0.6569 vs. 0.72) and LOTVS-CAP's 98.8\% (0.7406 vs. 0.75). Temporal metrics (mTTA; TTA@50) show <8.6\% variance across regimes vs. DRIVE's >27.18\% mTTA fluctuations. The causal strategy ensures stability by preserving precursors while suppressing spurious correlations, making CAMERA as first model optimizing both performance maxima and data-efficiency minima. 

\begin{table}[h]
    \centering
    \vspace{-5pt}
    \renewcommand{\arraystretch}{1.2}
    \resizebox{0.41\textwidth}{!}{
    \begin{tabular}{c|c|c|c|c}
    \hline
         & Training Set & AP & mTTA & TTA@50 \\
    \hline
        \multirow{3}{*}{DRIVE \citep{bao2021drive}} 
        & 50\%  & 0.6569  & 3.1242  & 3.028  \\
        & 75\%  & \underline{0.7173}  & \underline{3.6732}  & \textbf{3.6776}  \\
        & 100\% & \textbf{0.72}    & \textbf{4.2905}   & \underline{3.657}  \\
    \hline
        \multirow{3}{*}{LOTVS-CAP \citep{li2024cognitive}} 
        & 50\%  & 0.7406  & \underline{4.1038}  & 3.4494  \\
        & 75\%  & \textbf{0.7582}  & 3.9944  & \underline{3.5723}  \\
        & 100\% & \underline{0.75}    & \textbf{4.4643}  & \textbf{3.9512}  \\
    \hline
        \multirow{3}{*}{CAMERA} 
        & 50\%  & 0.7413  & 4.0987  & 3.9852  \\
        & 75\%  & \underline{0.7642}  & \underline{4.2823}  & \textbf{4.3687}  \\
        & 100\% & \textbf{0.8052}  & \textbf{4.4837}  & \underline{4.0583}  \\
    \hline
    \end{tabular}
    }
    \caption{Performance comparison under different training data proportions. \textbf{Bold} and \underline{underlined} denote top and second-best results. }
    \label{tab:my_label3}
    \vspace{-20pt}
\end{table}

The framework combining partial training with simulated test-time data loss addresses deployment challenges in accident anticipation systems. By limiting training data and introducing controlled observation gaps at testing, this method examines whether models develop essential causal accident relationships versus superficial dataset artifacts. Progressive sensor deterioration (via incremental missing ratios) forces prioritization of invariant spatio-temporal risk patterns over transient cues, mirroring real-world data scarcity and partial observability. This dual-stress paradigm reveals models' temporal sensitivity maintenance under information deficits while exposing fragile correlation dependencies collapsing in imperfect sensing environments. The methodology establishes benchmarks for robustness against real-time safety uncertainties, requiring reliable predictions despite incomplete data and dynamic occlusions.

Table~\ref{tab:my_label4} highlights CAMERA's robust accident anticipation under limited training data and simulated sensor failures. With 75\% and 50\% missing observations, CAMERA achieves 0.7449 AP - surpassing LOTVS-CAP (0.7387) by 0.84\% and DRIVE (0.6054) by 23.04\%. Its temporal superiority persists at 50\% training data under 20\% missing ratio: 4.0957s mTTA and 3.9947s TTA@50 exceed LOTVS-CAP (4.0655s/3.4969s) by up to 0.5s - critical for accident anticipation. The gap expands under extreme conditions (50\% training data, 50\% missing ratio): CAMERA maintains 3.6421s TTA@50 versus LOTVS-CAP's 3.4992s (+4.08\%) while DRIVE degrades to 2.9623s. CAMERA's minimal AP decay (0.7951→0.7308, -8.08\%) versus LOTVS-CAP's -3.22\% (0.7508→0.7266) under halved data/observations confirms its learning of invariant precursors over dataset artifacts. This resilience to partial observability and data scarcity makes CAMERA a paradigm shift in proactive safety systems.
\begin{table}[h]
    \centering
    \vspace{-5pt}
    \renewcommand{\arraystretch}{1.2} 
    \resizebox{1\linewidth}{!}{
    \begin{tabular}{c|c|c|c|c|c}
    \hline
    Model & Limited Dataset & Drop Rate & AP & mTTA & TTA@50 \\
    \hline
    \multirow{6}{*}{DRIVE \citep{bao2021drive}} 
     & \multirow{3}{*}{75\%} 
     & 10\% & \textbf{0.6551} & \underline{3.0913} & \underline{3.0828} \\
     &  & 20\% & \underline{0.6314} & \textbf{3.1087} & \textbf{3.1068} \\
     &  & 50\% & 0.6054 & 3.0687 & 3.0813 \\
     \cline{2-6}
     & \multirow{3}{*}{50\%} 
     & 10\% & \textbf{0.6259} & \textbf{3.1022} & \underline{2.9966} \\
     &  & 20\% & \underline{0.6128} & \underline{3.0928} & \textbf{3.0052} \\
     &  & 50\% & 0.5861 & 3.0467 & 2.9623 \\
    \hline
    \multirow{6}{*}{LOTVS-CAP \citep{li2024cognitive}} 
     & \multirow{3}{*}{75\%} 
     & 10\% & \textbf{0.7508} & \underline{4.0363} & \underline{3.4234} \\
     &  & 20\% & \underline{0.7506} & 4.0256 & \textbf{3.3703} \\
     &  & 50\% & 0.7387 & \textbf{4.2064} & 3.3083 \\
     \cline{2-6}
     & \multirow{3}{*}{50\%} 
     & 10\% & \textbf{0.7437} & \underline{4.0971} & \underline{3.4907} \\
     &  & 20\% & \underline{0.7410} & 4.0655 & 3.4969 \\
     &  & 50\% & 0.7266 & \textbf{4.1509} & \textbf{3.4992} \\
    \hline
    \multirow{6}{*}{CAMERA} 
     & \multirow{3}{*}{75\%} 
     & 10\% & \textbf{0.7951} & \textbf{4.4468} & \textbf{4.3952} \\
     &  & 20\% & \underline{0.7812} & \underline{4.3784} & \underline{4.1513} \\
     &  & 50\% & 0.7449 & 4.1786 & 3.7904 \\
     \cline{2-6}
     & \multirow{3}{*}{50\%} 
     & 10\% & \textbf{0.7895} & \textbf{4.2617} & \textbf{4.0866} \\
     &  & 20\% & \underline{0.7790} & \underline{4.0957} & \underline{3.9947} \\
     &  & 50\% & 0.7308 & 4.0995 & 3.6421 \\
    \hline
    \end{tabular}}
    \caption{Performance under varying drop rates for limited-dataset models. \textbf{Bold} and \underline{underlined} denote top and second-best results.}
    \label{tab:my_label4}
    \vspace{-15pt}
\end{table}

The MFE module constitutes CAMERA's perceptual backbone - its ablation (Model A in Table~\ref{tab:my_label5}) causes an 18.4\% AP drop (0.6571 vs. 0.8052), highlighting its critical role in processing heterogeneous inputs. Through dynamically calibrating RGB channels, MFE selectively amplifies obscured traffic agents under varying illumination while suppressing background noise. Simultaneously, it converts textual descriptions ("a motorbike runs red light") into semantic embeddings that anchor visual interpretations to traffic contexts. The dual-attention mechanism operating on driver gaze maps sharpens spatial focus by aligning machine attention with human perceptual patterns. Absent MFE's modality-specific processing, cross-modal alignment deteriorates, degrading mTTA by 1.2 seconds.

The AHF module bridges modality-specific representations - removal (Model B, Table~\ref{tab:my_label5}) degrades AP by 4.5\% (0.7689 vs. 0.8052), confirming its complementary synthesis. Through multi-scale harmonization and cross-modal co-activation, AHF weights visual, textual, and attentional cues to resolve scenario ambiguities. HSFA preserves spatial details while integrating global context for risks like lane intrusions. Cross-modal layers link traffic descriptions to visual patterns (e.g., "dodge action" with motion artifacts). BiBA's bifurcated projection ensures task-specific purity by separating spatial risks for heatmaps while retaining contextual semantics for classification. Overall, the AHF module is essential for cross-modal integration, enabling CAMERA to interpret complex scenarios. 

Bi-GRU functions as the temporal reasoning engine - exclusion (Model C, Table 5) decreases AP by 10.4\% (0.7214 vs. 0.8052) and mTTA by 22.7\% (3.46s vs. 4.48s), confirming its critical role in modeling accident precursors. GRU gates maintain long-range dependencies within the 3-5s prediction window to capture evolving risks like vehicle acceleration or erratic pedestrian paths. Bidirectional processing jointly analyzes forward-evolving threats (e.g., crossing motorbikes) and backward-traceable precursors (e.g., leading vehicle stops), establishing causal links between transient events and accident probabilities. The AMB mechanism enhances temporal coherence by filtering noise while preserving motion cues, contrasting with frame-wise models. This temporal integration improves risk probability curve reliability, suppressing false alarms during stable driving yet retaining abrupt hazard sensitivity. GRU hidden states form spatial risk localization foundations, ensuring heatmap consistency with temporal escalation patterns.
\begin{table}
    \centering
    \vspace{-5pt}
    \resizebox{0.48\textwidth}{!}{
    \begin{tabular}{c|c|c|c|c|c|c}
        \hline
        \cline{2-7}
         & \multicolumn{3}{c|}{Module} & \multicolumn{3}{c}{Metrics} \\
        \cline{2-7}
        Model     & MFE & AHF & Bi-GRU & AP & mTTA & TTA@R50 \\
        \hline
        A & \( \times \) & \( \bullet \) & \( \bullet \)  & 0.6571 & 3.2601 & 3.3874 \\
        B & \( \bullet \) & \( \times \) & \( \bullet \) & 0.7689 & 3.6515 & 3.4290 \\
        C & \( \bullet \) & \( \bullet \) & \( \times \) & 0.7214 & 3.4635 & 3.6052\\
        CAMERA     & \( \bullet \) & \( \bullet \) & \( \bullet \) & \textbf{0.8052} & \textbf{4.4837} & \textbf{4.0583} \\
        \hline
    \end{tabular}
    }
    \caption{Ablation results on the DADA-2000, where MFE donates Multi-Modal Feature Extraction, AHF donates Adaptive Hierarchical Fusion. \textbf{Bold} denotes the result of the complete CAMERA.}
    \label{tab:my_label5}
\end{table}

\subsection{Visualization}
\begin{figure}
    \centering
      \vspace{-10pt}
    \includegraphics[width=1\linewidth]{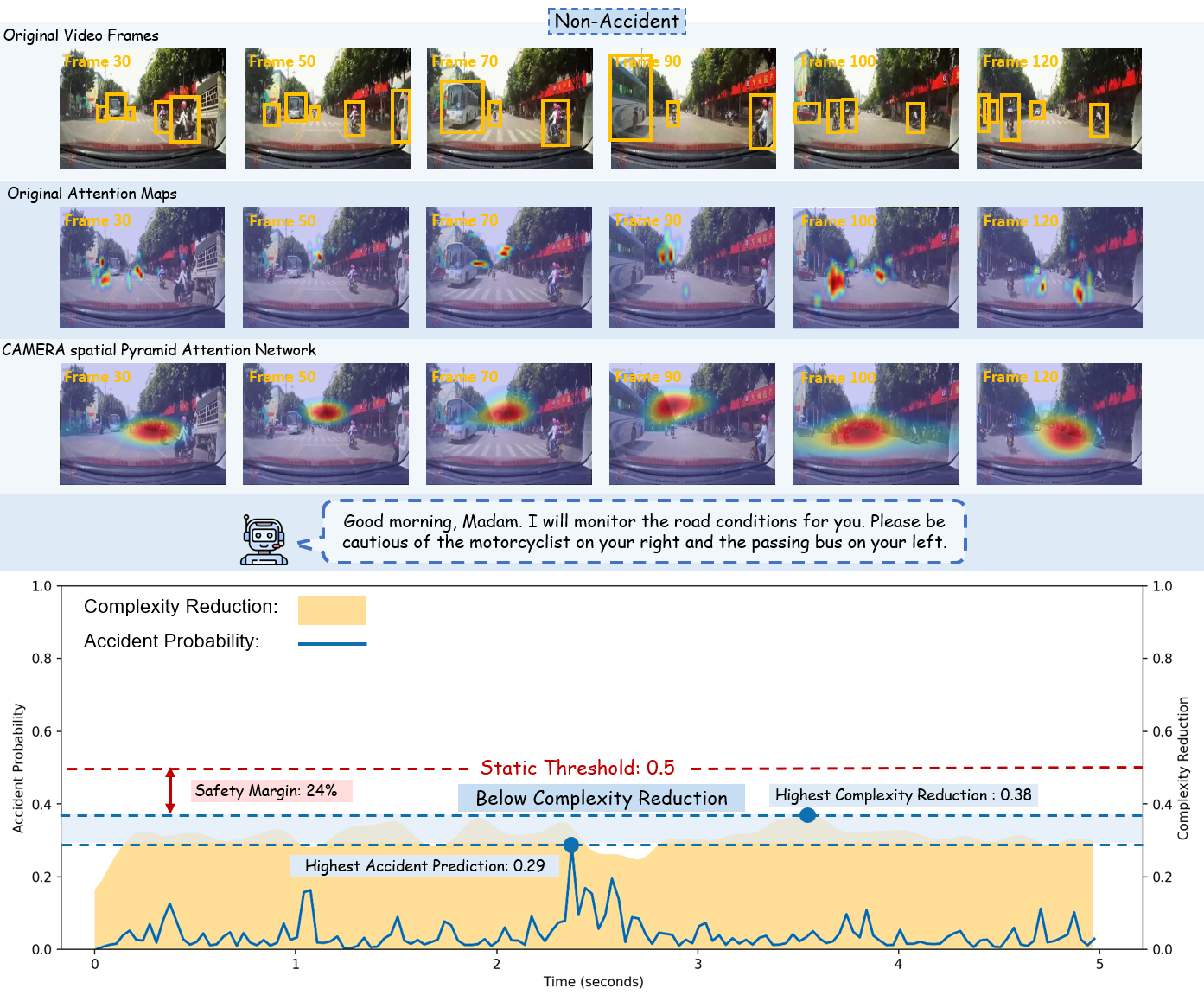}
      \vspace{-15pt}
    \caption{DADA-2000 non-accident analysis: (a) Detected agents (yellow); (b) Geo-Context VLM; (c) Dynamic thresholds.}
    \label{fig:enter-label2}
    \vspace{-4pt}
\end{figure}

\begin{figure}
    \centering
    \vspace{-8pt}
    \includegraphics[width=1\linewidth]{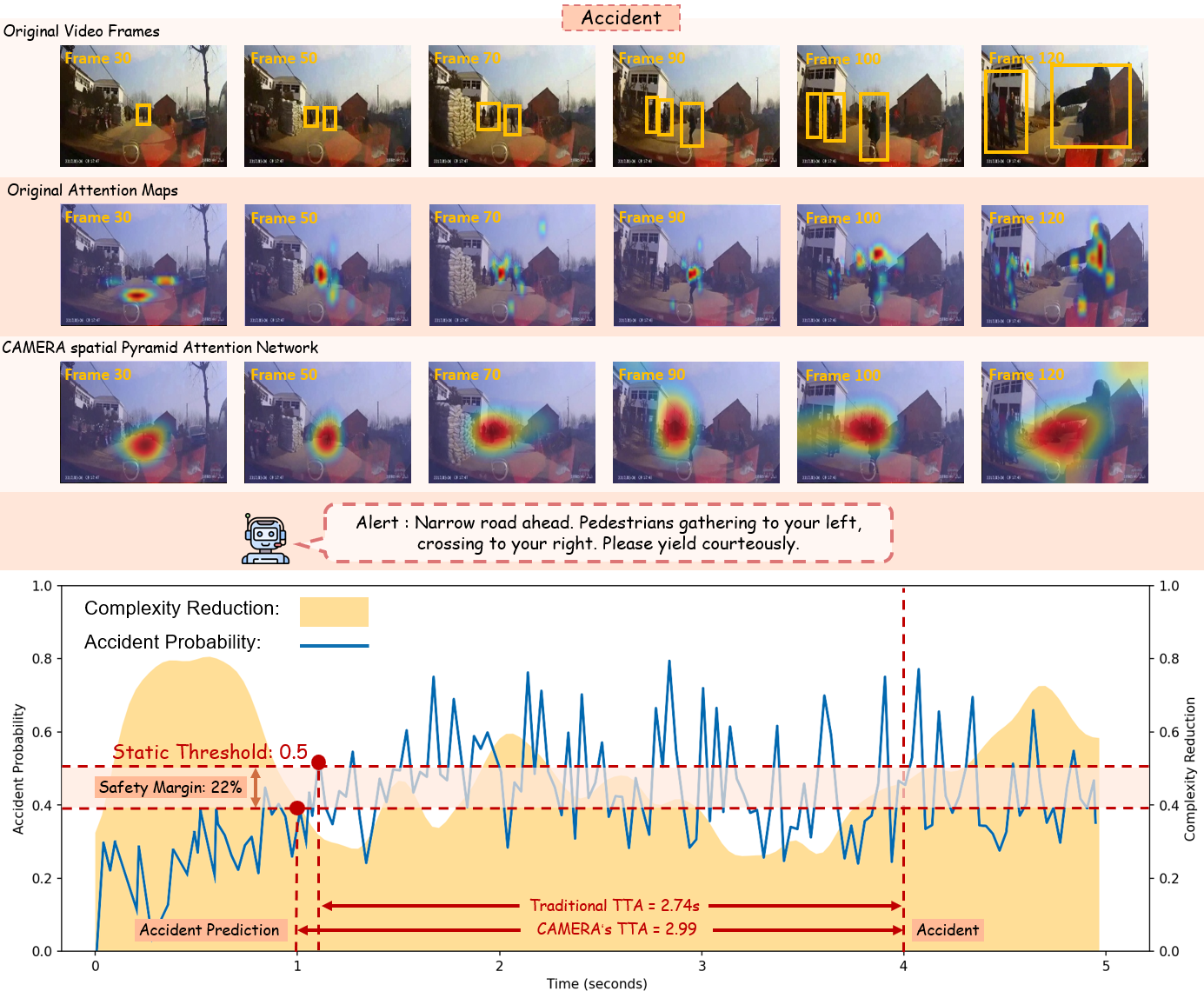}
      \vspace{-15pt}
    \caption{DADA-2000 accident analysis: (a) Detected agents (yellow); (b) Geo-Context VLM; (c) Dynamic thresholds.}
    \label{fig:enter-label3}
     \vspace{-2pt}
\end{figure}

We analyze accident and non-accident states separately (Fig.~\ref{fig:enter-label2} and Fig.~\ref{fig:enter-label3}). In the non-accident case (Fig.~\ref{fig:enter-label2}), the ego-vehicle travels on a bidirectional two-lane road without physical barriers, with double-yellow lines separating traffic and narrow motorcycle/bicycle lanes. Simultaneous interactions occur: two motorcycles maintain <0.5m lateral proximity on the right, and an opposing bus is present at 30° azimuth with a +15 km/h speed differential. CAMERA’s dual-metric framework identifies key dynamics: accident probability remains below complexity reduction, a dynamic threshold adapting to road geometry, traffic density, and kinematics. Complexity reduction decreases with spatial constraints and heterogeneous interactions, unlike static thresholds (e.g., 0.5), enhancing sensitivity. Geo-Context VLM links the bus’s trajectory (30° azimuth, 1.2 m/s² deceleration) and motorcycles’ encroachment (<0.5m deviation, 2.8 m/s lateral speed) to complexity reduction adjustments. This calibration aligns threshold sensitivity with kinematic threats, unlike a fixed 0.5 threshold. Such adaptability is crucial in mixed traffic, where static parameters fail to capture emerging risks from bidirectional interactions, variable lanes, and diverse behaviors.  

Fig.~\ref{fig:enter-label2} and Fig.~\ref{fig:enter-label3} employs a three-layer comparison to demonstrate CAMERA’s accuracy improvement through driver attention optimization: 1) raw pre-accident frames, 2) recorded driver attention heatmaps, and 3) multi-scale fused attention representations generated by CAMERA’s spatial pyramid architecture in MFE/AHF modules. In Figure~\ref{fig:enter-label3}, the pyramid network filters and amplifies fragmented attention patterns, prioritizing critical environmental agents. This optimizes resource allocation by focusing on salient regions, reducing redundancy. Geo-Context VLM enhances performance through context-aware hazard identification, producing alerts for road narrowing, left-side pedestrian clusters, and right-side crossings, with suggestions. Quantitative analysis shows CAMERA triggers warnings when predicted probability (~0.4) exceeds the dynamic threshold, achieving a 0.25-second advantage over static thresholds (0.5) while maintaining a 22\% safety margin. These results confirm CAMERA improves prognostic sensitivity, extending emergency response times and enabling accident avoidance.  

\begin{figure}
    \centering
    \vspace{-15pt}
    \includegraphics[width=1\linewidth]{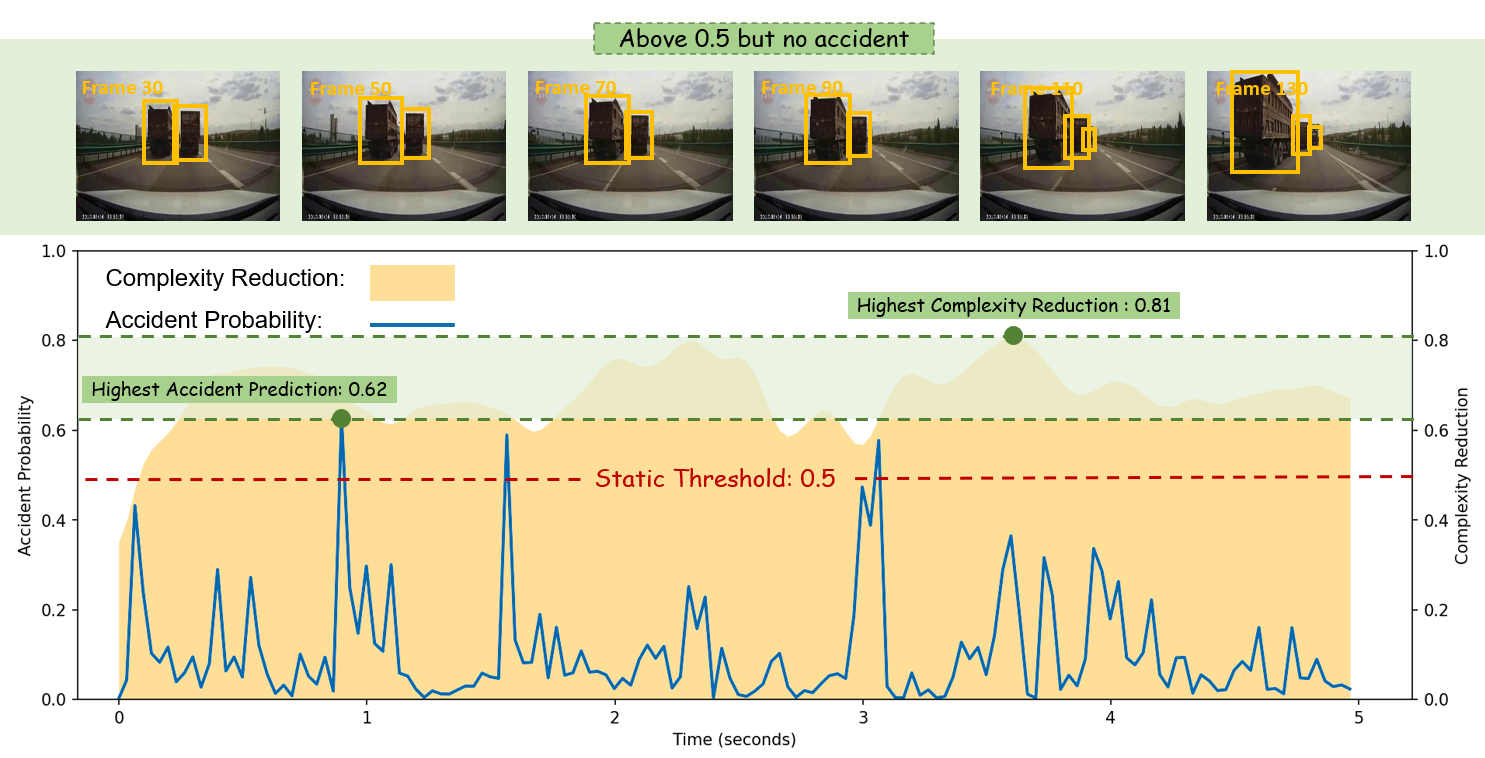}
  \vspace{-18pt}
    \caption{DADA-2000 conflict case: High-risk anticipation vs. null outcome, agents (yellow) and temporal data.}
    \label{fig:enter-label4}
    \vspace{-10pt}
\end{figure}

\begin{figure}
    \centering
    \includegraphics[width=1\linewidth]{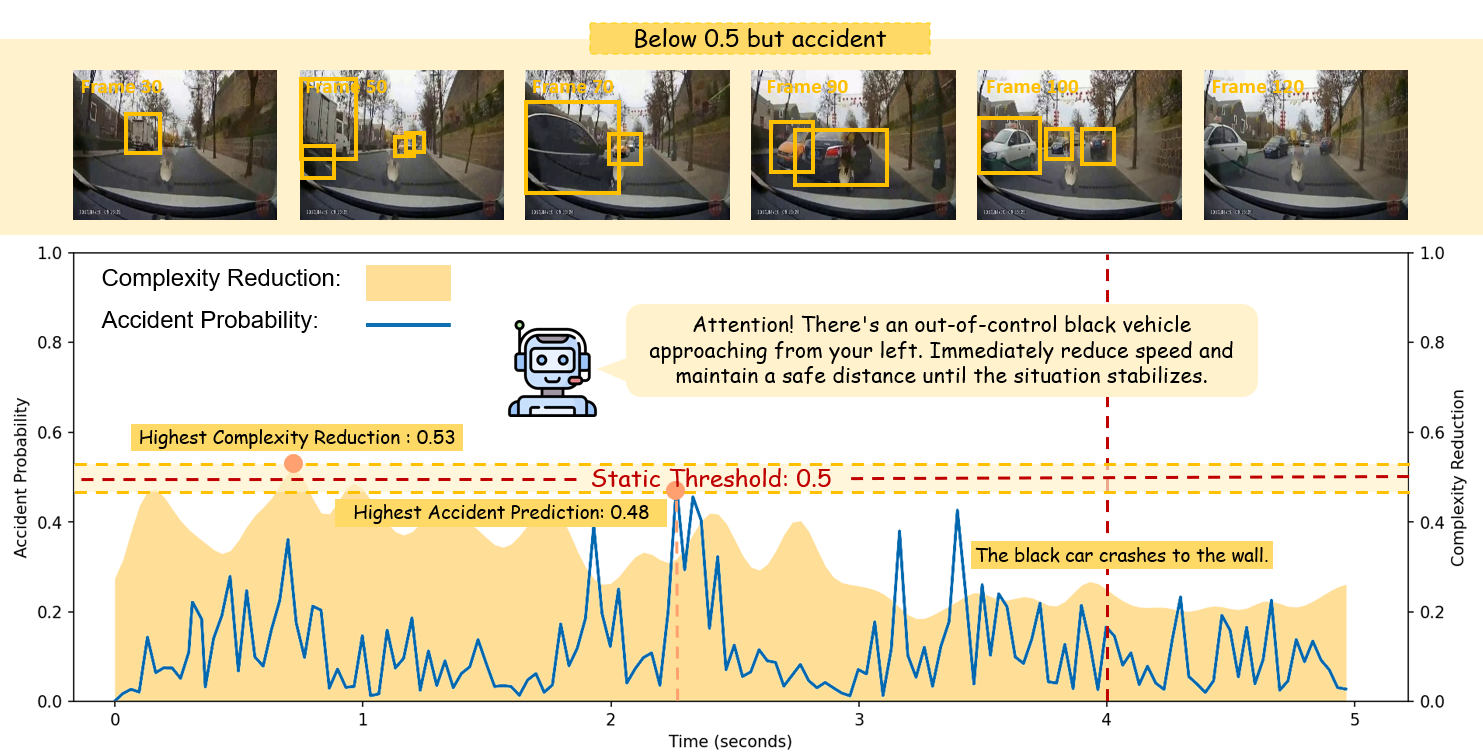}
        \vspace{-18pt}
    \caption{DADA-2000 conflict case: Low-risk anticipation vs. null outcome, agents (yellow) and temporal data.}
    \label{fig:enter-label5}

\end{figure}

The dynamic threshold based on environmental complexity reduction derives from its ability to adapt to evolving risks. A static 0.5 threshold may over-alarm in low-traffic scenarios by misclassifying moderate risks (e.g., accident probability = 0.5–0.7, Fig~\ref{fig:enter-label4}) or under-predict risks in multi-agent conflicts due to ignored interactions (probability 0.4–0.5 in high-complexity cases, Fig~\ref{fig:enter-label5}). The complexity-reduction metric adapts to real-time scene entropy, enhancing risk sensitivity calibration. As validated in Fig~\ref{fig:enter-label2}, the divergence between accident probability (0.0–0.29) and complexity reduction (lowered to 0.38) confirms CAMERA’s 24\% safety margin below the static 0.5 benchmark. By aligning thresholds with environmental stochasticity, the system achieves context-sensitive risk stratification—suppressing false alarms in trivial scenarios while detecting latent hazards, optimizing precision-timeliness trade-offs.

\section{Conclusion}
As autonomous driving advances toward full-scale commercialization, accurately detecting, localizing, and proactively warning of traffic risks is critical for safe navigation and fostering public trust in deployment. To address this, we propose CAMERA—a framework that enhances proactive traffic safety by aligning driver cognition with environmental dynamics via attention mapping, multimodal fusion, and context-aware thresholding. CAMERA achieves state-of-the-art performance (80.52\% AP, 89.64\% AUC) and enables human-centric risk communication. Its dynamic complexity reduction preserves a 24.00\% safety margin while enabling earlier warnings through real-time entropy adaptation. The VLM bridges machine perception and driver intuition by translating spatial risks into actionable alerts. Remaining challenges include computational demands and dependence on high-resolution gaze data; future work will explore lightweight models and semi-supervised learning to improve scalability and reduce annotation costs.

\section*{Acknowledgements}
This work was supported by the Science and Technology Development Fund of Macau [0122/2024/RIB2, 0215/2024/AGJ, 001/2024/SKL], the Research Services and Knowledge Transfer Office, University of Macau [SRG2023-00037-IOTSC, MYRG-GRG2024-00284-IOTSC], the Shenzhen-Hong Kong-Macau Science and Technology Program Category C [SGDX20230821095159012], the State Key Lab of Intelligent Transportation System [2024-B001], and the Jiangsu Provincial Science and Technology Program [BZ2024055].

\newpage
\bibliographystyle{ACM-Reference-Format}
\bibliography{references}
\end{document}